\documentstyle[12pt,axodraw,procsla]{article}

\catcode`\@=11
\def\@citexlow[#1]#2{\if@filesw\immediate\write\@auxout
	{\string\citation{#2}}\fi
\def\@citealow{}\@citelow{\@for\@citeblow:=#2\do
	{\@citealow\def\@citealow{,}\@ifundefined
	{b@\@citeblow}{{\bf ?}\@warning
	{Citation `\@citeblow' on page \thepage \space undefined}}
	{\csname b@\@citeblow\endcsname}}}{#1}}

\newif\if@cghi
\def\citelow{\@cghifalse\@ifnextchar [{\@tempswatrue
	\@citexlow}{\@tempswafalse\@citexlow[]}}
\def\@citelow#1#2{{#1\if@tempswa\typeout
	{IJCGA warning: optional citation argument
	ignored: `#2'} \fi}}
\catcode`\@=12

\renewcommand{\Im}{\mathop{\rm Im}\nolimits}
\def\slash#1{\setbox0=\hbox{$#1$}               
   \dimen0=\wd0                                 
   \setbox1=\hbox{/} \dimen1=\wd1               
   \ifdim\dimen0>\dimen1                        
      \rlap{\hbox to \dimen0{\hfil/\hfil}}      
      #1                                        
   \else                                        
      \rlap{\hbox to \dimen1{\hfil$#1$\hfil}}   
      /                                         
   \fi}                                         %

\begin{document}

\centerline{\normalsize\bf GAUGE INVARIANCE}
\centerline{\normalsize\bf IN}
\centerline{\normalsize\bf BOSON PRODUCTION}

\centerline{\footnotesize ROBIN G. STUART}
\baselineskip=13pt
\centerline{\footnotesize\it Randall Laboratory of Physics,
                             University of Michigan,
500 E. University}
\baselineskip=12pt
\centerline{\footnotesize\it Ann Arbor, Michigan 48109-1120, USA}
\centerline{\footnotesize E-mail: stuartr@umich.edu}

\vspace*{0.9cm}
\abstracts{
It is shown how to construct exactly gauge-invariant $S$-matrix elements
for processes involving unstable gauge particles such as the $W$ and
$Z^0$ bosons. The results are
applied to derive a physically meaningful expression
for the cross-section $\sigma(e^+e^-\rightarrow Z^0Z^0)$ and thereby
provide a solution to the long-standing
{\it problem of the unstable particle}. The problem of maintaining
QED gauge-invariance in the process
$e^+e^-\rightarrow\bar\nu_e e^-W^+\rightarrow\bar\nu_e e^-u\bar d$
is examined.}

\normalsize\baselineskip=15pt
\setcounter{footnote}{0}
\renewcommand{\thefootnote}{\alph{footnote}}

\section{Introduction}

Pair production of $W$ bosons has now been observed at LEP and both
$W$'s and $Z^0$'s will be copiously produced at upcoming colliders.
Yet both
the $W$ and $Z^0$ are unstable with quite sizable widths that cannot
be treated using the standard methods of $S$-matrix theory because of
the classic {\sl problem of the unstable particle}\cite{Peierls}.
Roughly stated the problem is as follows:
$S$-matrix theory deals with asymptotic in-states and out-states that
propagate from and to infinity. Unstable particles cannot exist as
asymptotic states because they decay a finite distance from the
interaction region. Indeed it is known\cite{Veltman} that the $S$-matrix
is unitary on the Hilbert space spanned by stable particle states and
hence there is not even any room to accommodate unstable particles as
external states. How can one use the $S$-matrix to calculate, say, the
production cross-section for an unstable particle when it cannot exist as
an asymptotic state?

In order to account for the finite width of the $W$ and $Z^0$ one
normally performs a Dyson summation of the one particle irreducible (1PI)
self-energy diagrams that effectively replaces the tree-level propagator
by a dressed propagator. However the 1PI self-energy
is gauge-dependent starting at ${\cal O}(\alpha)$ and hence
dressed propagator is gauge-dependent at all orders in $\alpha$.
If the dressed propagator is used in a finite-order calculation the
resulting matrix element will also be gauge-dependent at some order.
This gauge-dependence should be
viewed as an indicator that the approximation scheme being used is
inconsistent or does not represent a physical observable and is a
latter-day manifestation of the problem of the unstable particle.

\section{The Problem of the Unstable Particle}

In this section we obtain a physical and therefore gauge-invariant
property that may be used to identify an unstable particle without
requiring that it appear in the final state.

Recall that the coordinate space dressed propagator for a scalar
particle has an integral representation
\begin{equation}
\Delta(x^\prime-x)=\int \frac{d^4k}{(2\pi)^4}
                   \frac{e^{-ik\cdot(x^\prime-x)}}
                                 {k^2-m^2-\Pi(k^2)+i\epsilon}
\label{eq:dressedprop}
\end{equation}
The integrand has a pole at $k^2=s_p$ where $s_p$ is a solution of the
equation $s-m^2+\Pi(s)=0$ and an analytic function,
$F(s)$, via the relation $s-m^2+\Pi(s)=(s-s_p)/F(s)$.
The dressed propagator can then be written as
\begin{equation}
\Delta(x^\prime-x)=\int \frac{d^4k}{(2\pi)^4}
        e^{-ik\cdot(x^\prime-x)}
         \left[\frac{F(s_p)}{k^2-s_p}
         +\frac{F(k^2)-F(s_p)}{k^2-s_p}\right]
\label{eq:splitprop}
\end{equation}
that separates resonant and non-resonant pieces. Performing the
$k_0$ integration on the resonant part gives
\begin{eqnarray}
& &\Delta(x^\prime-x)=-i\int \frac{d^3k}{(2\pi)^3 2k_0}
         e^{-ik\cdot(x^\prime-x)}\theta(t^\prime-t)F(s_p)\nonumber\\
      & &\qquad+\int\frac{d^4k}{(2\pi)^4}\frac{F(k^2)-F(s_p)}{k^2-s_p}
         -i\int \frac{d^3k}{(2\pi)^3 2k_0}
         e^{ik\cdot(x^\prime-x)}\theta(t-t^\prime)F(s_p)\nonumber
\end{eqnarray}
where $k_0=\sqrt{\vec{k}^2+s_p}$.
The non-resonant term contributes only for $t=t^\prime$ and so
represents a contact interaction. The resonant part spits into two terms
that contribute when $t>t^\prime$ or $t<t^\prime$ and
therefore connects points $x$ and $x^\prime$ that are
separated by a finite distance in space-time.
When a similar analysis is applied to a physical matrix
element one concludes that the resonant
part describes a process in which there is a finite space-time separation
between the initial-state vertex, $V_i$, and the final-state vertex, $V_f$.
In other words, the resonant term describes the finite propagation of a
physical unstable particle. The non-resonant background represents prompt
production of the final state. As these two possibilities are, in principle,
physically distinguishable, they must be separately gauge-invariant.

We can thus use finite propagation as a tag for identifying unstable
particles without requiring that they appear in the final state.
This is, after all, the way $b$-quarks are identified in vertex detectors.
A production cross-section for an unstable particle is obtained by
extracting the resonant part of the matrix element for a process containing
that particle in an intermediate state and summing over all possible
decay modes.

\section{The process $\lowercase{e^+e^-}\rightarrow Z^0Z^0$}

In this section we will calculate the cross-section for
$e^+e^-\rightarrow Z^0Z^0$. This is of both theoretical and practical
importance. On the theoretical side it represents an example of a
calculation of the production cross-section for unstable particles.
On the practical side, at high energies $e^+e^-\rightarrow Z^0Z^0$
will be a dominant source of fermion pairs $(f_1\bar f_1)$ and
$(f_2\bar f_2)$ due to its double resonant enhancement and hence
$\sigma(e^+e^-\rightarrow Z^0Z^0)$ is an excellent approximation
to the cross-section for 4-fermion pair production.
If the experimental situation warrants it, background terms
can also be included without difficulty.
A more detailed account can be found in ref.~\citelow{Unstable}.

In the case of $e^+e^-\rightarrow f\bar f$, dealt with in
ref.s~\citelow{Stuart1,Stuart3} the invariant mass squared of the $Z^0$,
$s$, is fixed by the momenta of the
incoming $e^+e^-$. For the process $e^+e^-\rightarrow Z^0Z^0$ the invariant
mass of the produced $Z^0$'s is not constant
and must be somehow included in phase
space integrations. It is not immediately clear how to do this and without
further guidance from $S$-matrix theory there would seem to be considerable
flexibility in how to proceed. A new ingredient is required and that is
to realize that an expression for an $S$-matrix element can always be
divided into a part that is a Lorentz-invariant function of the kinematic
invariants of the problem and Lorentz-covariant objects, such as $\slash{p}$
etc. The latter are known as {\it standard covariants\/}\cite{Hearn,Hepp1,Williams,Hepp2}.
It is the Lorentz invariant part that
satisfies the requirements of analytic
$S$-matrix theory and from which the resonant
and non-resonant background parts are extracted while the Lorentz
covariant part is untouched.

To calculate $\sigma(e^+e^-\rightarrow Z^0Z^0)$, we begin by constructing
the cross-section for
$e^+e^-\rightarrow Z^0Z^0\rightarrow (f_1\bar f_1)(f_2\bar f_2)$,
and will eventually sum over all fermion species.
The squared invariant masses of the $f_1\bar f_1$ and $f_2\bar f_2$ pairs
will be denoted $p_1^2$ and $p_2^2$.

To obtain the piece of the matrix element that corresponds to finite
propagation of both $Z^0$'s we extract the leading term in a Laurent
expansion in $p_1^2$ and $p_2^2$ of the analytic Lorentz-invariant
part of matrix element leaving the Lorentz-covariant part
untouched. This is the doubly-resonant term and is given by
\begin{eqnarray}
{\cal M}&=&\sum_i [\bar v_{e^+} T^i_{\mu\nu} u_{e^-}]
M_i(t,u,s_p,s_p)\nonumber\\
& &\ \ \ \ \ \ \ \times\frac{F_{ZZ}(s_p)}{p_1^2-s_p}
[\bar u_{f_1}\gamma^\mu(V_{Zf_L}(s_p)\gamma_L
                       +V_{Zf_R}(s_p)\gamma_R) v_{\bar f_1}]
\label{eq:resZZ}\\
& &\ \ \ \ \ \ \ \times\frac{F_{ZZ}(s_p)}{p_2^2-s_p}
[\bar u_{f_2}\gamma^\nu(V_{Zf_L}(s_p)\gamma_L
                       +V_{Zf_R}(s_p)\gamma_R) v_{\bar f_2}]
\nonumber
\end{eqnarray}
where $T_{\mu\nu}^i$ are Lorentz covariant tensors that span the tensor
structure of the matrix element and $\gamma_L$, $\gamma_R$ are the
usual helicity projection operators.
The $M_i$, $\Pi_{ZZ}$ and $V_{Zf}$
are Lorentz scalars that are analytic functions of the
independent kinematic Lorentz invariants of the problem.
$F_{ZZ}$ is defined in the way described following
eq.(\ref{eq:dressedprop}).
It should be emphasized that eq.(\ref{eq:resZZ}) is the exact form
of the doubly-resonant matrix element to all orders in perturbation
theory that we will now specialize to leading order.

In the procedure described above, one starts by extracting the resonant
term in a scattering amplitude by Laurent expansion about the exact pole
position $s_p$ and then specializes to lower orders by further expanding
about the renormalized mass. Other authors\cite{AeppCuypOlde,AeppOldeWyl}
have attempted to apply the techniques described above by first expanding
about the renormalized mass and then adding a finite width in the denominator
of the resonant part by hand. That procedure cannot be justified and leads
to problems when one treats processes like $e^+e^-\rightarrow W^+W^-$.
It gives rise to spurious {\it threshold singularities} or complex
scattering angles due the production threshold's branch point
being incorrectly located on the real axis.

It is important to note that Laurent expansion refers to how the
resonant term is
identified. The aim is only to separate ${\cal M}$ into gauge-invariant
resonant and background pieces and there is no necessity to perform
a Laurent expansion beyond its leading term although this can provide
a useful way for parameterizing electroweak data\cite{SMatData}.
Although in what follows the background terms are systematically dropped,
they can be kept if the experimental situation should require it.

In lowest order eq.(\ref{eq:resZZ}) becomes, up to overall multiplicative
factors,
\begin{eqnarray}
{\cal M}&=&\sum_{i=1}^2 [\bar v_{e^+} T^i_{\mu\nu} u_{e^-}]M_i\nonumber\\
& &\ \ \ \ \times\frac{1}{p_1^2-s_p}
[\bar u_{f_1}\gamma^\mu(V_{Zf_L}\gamma_L+V_{Zf_R}\gamma_R)v_{\bar f_1}]
\label{eq:lowestZZ}\\
& &\ \ \ \ \times\frac{1}{p_2^2-s_p}
[\bar u_{f_2}\gamma^\nu(V_{Zf_L}\gamma_L+V_{Zf_R}\gamma_R)v_{\bar f_2}].
\nonumber
\end{eqnarray}
where
$T^1_{\mu\nu}=\gamma_\mu(\slash{p}_{e^-}-\slash{p}_1)\gamma_\nu$,
$M_1=t^{-1}$;
$T^2_{\mu\nu}=\gamma_\nu(\slash{p}_{e^-}-\slash{p}_2)\gamma_\mu$,
$M_2=u^{-1}$
and the final state vertex corrections take the form
$V_{Zf_L}=ie\beta_L^f\gamma_L$ and $V_{Zf_R}=ie\beta_R^f\gamma_R$,
The left- and right-handed couplings of the $Z^0$ to a fermion $f$ are
\[
\beta_L^f=\frac{t_3^f-\sin^2\theta_W Q^f}{\sin\theta_W\cos\theta_W},
\hbox to 2cm{}
\beta_R^f=-\frac{\sin\theta_W Q^f}{\cos\theta_W}.
\]

Squaring the matrix element and integrating over the final state
momenta for fixed $p_1^2$ and $p_2^2$ gives
\begin{equation}
\frac{\partial^3\sigma}{\partial t\,\partial p_1^2\,\partial p_2^2}
                   =\frac{\pi\alpha^2}{s^2}
                    (\vert\beta_L^e\vert^4+\vert\beta_R^e\vert^4)
                    \rho(p_1^2)\ \rho(p_2^2)
                   \left\{\frac{t}{u}+\frac{u}{t}
                     +\frac{2(p_1^2+p_2^2)^2}{ut}
                     -p_1^2p_2^2\left(\frac{1}{t^2}+\frac{1}{u^2}\right)
                     \right\}
\label{eq:diffxsec}
\end{equation}
with
\begin{eqnarray*}
\rho(p^2)&=&\frac{\alpha}{6\pi}
    \sum_f(\vert\beta_L^f\vert^2+\vert\beta_R^f\vert^2)
          \frac{p^2}{\vert p^2-s_p\vert^2}
               \theta(p_0)\theta(p^2)\\
 &\approx&\frac{1}{\pi}
         .\frac{p^2 (\Gamma_Z/M_Z)}{(p^2-M_Z^2)^2+\Gamma_Z^2 M_Z^2}
          \theta(p_0)\theta(p^2)
\end{eqnarray*}
where the sum is over fermion species.
Note that $\rho(p^2)\rightarrow \delta(p^2-M_Z^2)\theta(p_0)$
as $\Im(s_p)\rightarrow 0$ which is the result obtained by cutting a
free propagator. The variables $s$, $t$, $u$, $p_1^2$ and $p_2^2$ in
eq.(\ref{eq:diffxsec}) arise from products of standard covariants and external
wave functions and therefore take real values dictated by the kinematics.

Integrating over $t$, $p_1^2$ and $p_2^2$ leads to
\begin{equation}
\sigma(s)=\int_0^s dp_1^2
          \int_0^{(\sqrt{s}-\sqrt{p_1^2})^2} dp_2^2
          \sigma(s;p_1^2,p_2^2)\ \rho(p_1^2)\ \rho(p_2^2),
\label{eq:ZZxsec}
\end{equation}
where
\begin{eqnarray*}
\sigma(s;p_1^2,p_2^2)&=&\frac{2\pi\alpha^2}{s^2}
       (\vert\beta_L^e\vert^4+\vert\beta_R^e\vert^4)\\
&\times&
       \left\{\left(\frac{1+(p_1^2+p_2^2)^2/s^2}
{1-(p_1^2+p_2^2)/s}\right)\ln\left(\frac{-s+p_1^2+p_2^2+\lambda}
                                    {-s+p_1^2+p_2^2-\lambda}\right)
-\frac{\lambda}{s}\right\}
\end{eqnarray*}
and $\lambda=\sqrt{s^2+p_1^4+p_2^4-2sp_1^2-2sp_2^2-2p_1^2p_2^2}$.
For $p_1^2=p_2^2=M_Z^2$ this agrees with known results\cite{Brown}.

\section{The Process $e^+e^-\rightarrow\bar\nu_e e^-u\bar d$}

Massless particles, such as the photon, are always problematic for
analytic $S$-matrix analyses. In this section the process
$e^+e^-\rightarrow\bar\nu_e e^-u\bar d$
will be considered with particular emphasis on how take into account
the finite width of the $W$ boson while at the same time maintaining
QED gauge invariance. The dominant contribution to this process is
expected to come from physical $W^+$ production in which
$e^+e^-\rightarrow\bar\nu_e e^-W^+\rightarrow\bar\nu_e e^-u\bar d$
via diagrams of the type shown in Fig.1.
Assume for the moment that the $W$ is stable and therefore has the usual
propagator $\sim(p^2+M_W^2)^{-1}$. It is expected that the matrix element
for this process should satisfy the usual condition for QED gauge-invariance
\begin{equation}
q_\mu G^\mu=0
\label{eq:gaugecondition}
\end{equation}
where $G^\mu$ is the matrix element with the external electromagnetic
current and photon propagator
amputated and $q_\mu$ is the momentum in the photon propagator.
When this test is applied to the diagrams of Fig.1.
one finds that (\ref{eq:gaugecondition}) does not hold. The reason is
that there exist other Feynman diagrams that produce the same final state
and only when these are included is the condition (\ref{eq:gaugecondition})
correctly satisfied\cite{CERNgroup}. However the additional diagrams that
must be included do not contain internal $W$'s that could go on-mass-shell
and are therefore expected to be subdominant.

\begin{figure}[htb]
\begin{center}
\begin{picture}(90,100)(0,0)
\ArrowLine(10,90)(35,80)
\ArrowLine(35,80)(80,90)
\ArrowLine(80,10)(35,20)
\ArrowLine(35,20)(10,10)
\ArrowLine(85,35)(70,50)
\ArrowLine(70,50)(85,65)
\Photon(35,80)(40,50){2}{4}
\Photon(40,50)(35,20){2}{4}
\Photon(40,50)(70,50){2}{4}
\Vertex(35,80){1.2}
\Vertex(35,20){1.2}
\Vertex(40,50){1.2}
\Vertex(70,50){1.2}
\put(08,90){\makebox(0,0)[r]{$e^-$}}
\put(08,10){\makebox(0,0)[r]{$e^+$}}
\put(82,90){\makebox(0,0)[l]{$e^-$}}
\put(82,10){\makebox(0,0)[l]{$\bar{\nu}_e$}}
\put(87,65){\makebox(0,0)[l]{$u$}}
\put(87,35){\makebox(0,0)[l]{$\bar{d}$}}
\put(34,65){\makebox(0,0)[r]{$\gamma$}}
\put(35,35){\makebox(0,0)[r]{$W$}}
\put(57,55){\makebox(0,0)[b]{$W^+$}}
\end{picture}
\qquad
\begin{picture}(90,90)(0,-10)
\ArrowLine(10,70)(35,60)
\ArrowLine(35,60)(80,70)
\ArrowLine(80, 0)(55,18)
\ArrowLine(55,18)(35,20)
\ArrowLine(35,20)(10,10)
\ArrowLine(85,25)(70,35)
\ArrowLine(70,35)(85,50)
\Photon(35,60)(35,20){2}{5}
\Photon(55,18)(70,35){2}{3}
\Vertex(35,60){1.2}
\Vertex(35,20){1.2}
\Vertex(55,18){1.2}
\Vertex(70,35){1.2}
\put(08,70){\makebox(0,0)[r]{$e^-$}}
\put(08,10){\makebox(0,0)[r]{$e^+$}}
\put(82,70){\makebox(0,0)[l]{$e^-$}}
\put(82, 0){\makebox(0,0)[l]{$\bar{\nu}_e$}}
\put(87,50){\makebox(0,0)[l]{$u$}}
\put(87,25){\makebox(0,0)[l]{$\bar{d}$}}
\put(30,40){\makebox(0,0)[r]{$\gamma$}}
\put(70,30){\makebox(0,0)[br]{$W^+$}}
\end{picture}
\\
\end{center}
\caption[]{Tree-level diagrams contributing to the resonant part of the process
           $e^+e^-\to\bar\nu_e e^-W^+\to\bar\nu_e e^-u\bar d$.
        {\small (courtesy of the authors of ref.\citelow{CERNgroup}) } }
\end{figure}
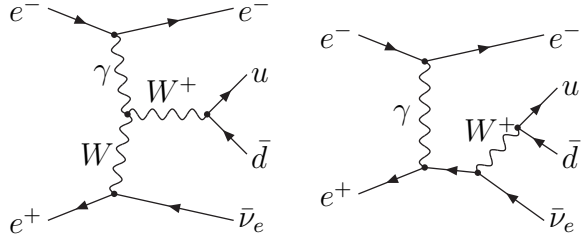

It might be thought that gauge dependence in subdominant terms of
no concern but for the process in question it plays a crucial r\^ole.
It can be shown\cite{CERNgroup}
that as $q^2\rightarrow 0$ the resulting cross-section behaves as
$\sim 1/q^2$ provided (\ref{eq:gaugecondition}) is satisfied and
$\sim 1/q^4$ when it is not. Without the maintenance of QED gauge
invariance the cross-section then blows up too strongly as
$q^2\rightarrow 0$.

The inclusion of subdominant diagrams to cancel the gauge-dependence
in the dominant contribution contrasts sharply with what was done in
section~3. The Laurent expansion extracts an exactly gauge-invariant
resonant term and eliminates from it any non-resonant (subdominant)
pieces. These pieces that are removed may be ultimately combined with
other contributions to yield an overall gauge-invariant non-resonant
contribution if required.

Now consider the case when the finite width of the $W$ boson in the
diagrams of Fig.1 is to be taken into account. If the procedure
described in sections~2 and 3 is applied to the process
$e^+e^-\rightarrow\bar\nu_e e^-W^+\rightarrow\bar\nu_e e^-u\bar d$
to extract the dominant resonant piece then the resulting matrix
element does not satisfy the condition (\ref{eq:gaugecondition}) and
the resulting cross-section blows up too strongly as
$q^2\rightarrow0$. A number
of solutions have been proposed and their relative merits examined in
ref.~\citelow{CERNgroup}. One that has received much attention is that of
Baur and Zeppenfeld\cite{BaurZeppen}. Their solution is to add the
imaginary parts of certain higher-order diagrams, specifically the
1-loop fermionic correction to the $W^+W^-\gamma$ vertex, which they
show exactly removes the QED gauge dependence. This method has the
practical disadvantage that it requires an additional calculation
that is significantly more complicated to obtain than the rest of the
tree-level amplitude. Moreover once it is calculated it is immediately
largely canceled by the gauge-dependent terms in the resonant part.

In this section we describe an alternative more economical approach
that lies more in line with the strategies adopted in section~3.
Instead of seeking additional higher-order gauge-dependent corrections
to cancel the QED gauge dependence, we eliminate it by means of a
projection operator under which the exact matrix element is known to
be invariant.

Suppose the approach of section~3 has been used to obtain the leading
resonant contribution to the process
$e^+e^-\rightarrow\bar\nu_e e^-W^+\rightarrow\bar\nu_e e^-u\bar d$.
After amputating the external electromagnetic current and photon
propagator a Green's function, $G_\mu$, is obtained that can be written
in terms of a set of standard covariants, $l_i^\mu$, and associated
Lorentz scalar functions, $A_i(p_1^2,p_2^2,..)$,
\begin{equation}
G^\mu=\sum_i l_i^\mu(p_1,p_2,...) A_i(p_1^2,p_2^2,...)
\label{eq:greenfunc}
\end{equation}
where the arguments $p_1$, $p_2$, etc.\ are 4-momenta that appear in the
problem.
The $l_i^\mu$, of course, must transform in the same way that $G^\mu$ does.

The gauge condition on Green's functions $q_\mu G^\mu=0$ means that the
$A_i$ are not linearly independent since they satisfy the condition
$\sum_i (q\cdot l_i)A_i=0$.

Gauge-invariant Green's
functions can be obtained using the methods of Bardeen and
Tung\cite{BardeenTung}.
Suppose that we have some exact Green's function expressed
in terms of standard covariants and Lorentz scalars as in
eq.(\ref{eq:greenfunc}). Since $q_\mu G^\mu=0$ the Green's function is
invariant under of the action of the operator
\begin{equation}
I^\mu_\nu=g^\mu_\nu-\frac{p^\mu q_\nu}{p\cdot q}
\label{eq:projoper}
\end{equation}
where $p$ is some conveniently chosen momentum.
The spinor-Lorentz tensor structure of the Green's function is spanned by
the set $\{I^\mu_\nu l_i^\nu\}$. That is
\begin{equation}
G^\mu(p_1,p_2,...)=\sum_i \left(I_\nu^\mu l_i^\nu(p_1,p_2,...)\right)
                   A_i(p_1^2,p_2^2,...).
\end{equation}
For an approximate Green's function that does not satisfy the
QED gauge-condition (\ref{eq:gaugecondition}), the operator
$I_\nu^\mu$ serves as a
projection operator onto $\{I^\mu_\nu l_i^\nu\}$. From the
$\{I^\mu_\nu l_i^\nu\}$ it is always possible
to construct a new basis that is free of
kinematic singularities that might have occurred when, for example,
$(p\cdot q)=0$. In practice, however, it is simpler to just choose
the vector $p$ such that $(p\cdot q)$ is not small in the kinematic
region of interest.

In general consider a Green's function $G^\mu$ calculated using some
incomplete expansion up to some order. It will consist of two parts
\begin{equation}
G^\mu=G_0^\mu+G_1^\mu
\end{equation}
where $G_0^\mu$ is a consistent gauge-invariant contribution
correct to given order of the calculation. $G_1^\mu$ is a spurious
higher-order gauge-dependent correction. Because $G_0^\mu$
satisfies the gauge condition, $q_\mu G_0^\mu=0$, it is invariant
under the action of the operator $I_\mu^\nu$.
Thus at a given order we may make the replacement
\[
G^\mu\rightarrow I^\mu_\nu G^\nu
             =G_0^\mu+I^\mu_\nu G_1^\nu.
\]
Only $G_1^\mu$ is affected by the projection operator but
since it is of higher order this is of no concern. The important point is
that it now satisfies the gauge condition $q_\mu(I^\mu_\nu G_1^\nu)=0$
and is prevented from causing too strong a numerical blow up as
$q^2\rightarrow0$.

Now, of course, the result depends on the choice of the vector, $p$,
which when judiciously done can be used to significantly simplify
calculations. The dependence on $p$, as it appears only in higher-orders
is analogous to renormalization scheme dependence, or the $\mu$-dependence
in the $\overline{\rm MS}$ scheme, that is an inescapable feature of any
perturbative expansion. As with scheme dependence, we can be sure that it
will become ever weaker as higher orders included since the exact Green's
function is invariant under the action of $I^\mu_\nu$.

Details of applying the projection operator to the process
$e^+e^-\rightarrow\bar\nu_e e^-W^+\rightarrow\bar\nu_e e^-u\bar d$
may be found in ref.~\citelow{SingleW} but may be summarized as follows:\\
The methods of section~3 are used to obtain the resonant part
of the matrix element ${\cal M}$ which takes the form
\begin{equation}
{\cal M}=J^\gamma_\lambda P_\mu^\lambda G^\mu
\end{equation}
where $J^\gamma_\lambda$ is the electromagnetic current of the electron
and $P_\mu^\lambda$ is the photon propagator.
It is simply a matter of inserting the projection operator,
(\ref{eq:projoper}), to obtain
\begin{equation}
{\cal M}=J^\gamma_\lambda P^\lambda_\nu(I^\nu_\mu G^\mu).
\end{equation}
The terms in parentheses satisfy the gauge-condition
(\ref{eq:gaugecondition}) and therefore the cross-section formed from
${\cal M}$ behaves $\sim 1/q^2$ as $q^2\rightarrow0$. Terms eliminated
by the projection operator must cancel in the exact matrix element.

\end{document}